# New critical phenomena in 2d quantum gravity

*Jan Ambjørn* and *Gudmar Thorleifsson*

The Niels Bohr Institute
Blegdamsvej 17, DK-2100 Copenhagen Ø, Denmark

*Mark Wexler*

Service de Physique Théorique de Saclay
91191 Gif-sur Yvette Cedex, France

## Abstract

We study $q = 10$ and $q = 200$ state Potts models on dynamical triangulated lattices and demonstrate that these models exhibit continuous phase transitions, contrary to the first order transition present on regular lattices. For $q = 10$ the transition seems to be of 2nd order, while it seems to be of 3rd order for $q = 200$. For $q = 200$ the phase transition also induces a transition between typical fractal structures of the piecewise linear surfaces corresponding to the triangulations. The typical surface changes from having a tree-like structure to a fractal structure characterizing pure gravity when the temperature drops below the critical temperature. An investigation of the alignment of spin clusters shows that they are strongly correlated to the underlying fractal structure of the triangulated surfaces.

# 1 Introduction

Important progress was made when it was understood that Euclidean quantum field theories can be described as the scaling limit of statistical models on hypercubic lattices. When a correlation length diverged at a critical point of the statistical model one could forget about the underlying lattice structure and define a continuum field theory. This point of view added many new aspects to our understanding of quantum field theory and especially to the concept of renormalization. Two-dimensional quantum gravity coupled to matter also allows a lattice regularization, where space-time, viewed as a two-dimensional manifold, is represented by a triangulation, i.e. approximated by a piecewise linear manifold [3, 2, 4]. Due to the dynamical nature of space-time in quantum gravity we cannot confine ourselves to a fixed lattice, but have to sum over all possible triangulations of a fixed topology.[1] Here we will only consider manifolds with spherical topology. The statistical model is now defined by specifying the matter interaction and the observables on an arbitrary triangulation and defining the partition function and average value of observables with respect to the annealed average over all triangulations.

As an example consider the $q$-states Potts model. On a regular two-dimensional lattice we can define it by

$$Z(\beta, N) = \sum_{\{q_i\}} \exp\left(\beta \sum_{(i,j)} \delta_{\sigma_i \sigma_j} - 1\right), \tag{1}$$

where $\sigma_i \in \{1, \ldots, q\}$, $i$ denotes a lattice site, $\{\sigma_i\}$ a spin configuration on the lattice and $\sum_{(i,j)}$ the summation over all neighboring pairs of lattice sites. Finally $N$ denotes the volume, which we take to be the number of lattice sites. Let us now couple the model to quantum gravity by the use of dynamical triangulations. Consider a triangulation $T$ of the sphere and place the Potts spins in the center of each triangle. The index $i$ in (1) now labels the triangles.[2] The neighbors of a given triangle $i$ are the triangles having a link in common with it. This defines $Z_T(\beta)$ for the given triangulation by formula (1), $N$ being the number of triangles in $T$. The annealed partition function for a finite volume $N$ is then defined by

$$Z(\beta, N) = \sum_{T \in \mathcal{T}(N)} Z_T(\beta). \tag{2}$$

where $\mathcal{T}$ denotes a suitable class of triangulations of the sphere. We will later discuss in some detail the precise choice of $\mathcal{T}$.

The $q$-state Potts models on regular two-dimensional lattices have a critical inverse temperature $\beta_c$. Below $\beta_c$ the spins are not aligned. Above $\beta_c$ the system is magnetized in the infinite volume limit. For $q = 2, 3$ and 4 the phase transition is second order and the associated central charge is $c = 1/2, 4/5$ and 1, respectively. The $q = 2$ (Ising) model can also be solved explicitly when coupled to gravity, i.e. the partition function (2) can be calculated by use of matrix model techniques. It is remarkable that although it is of course impossible to solve the models on some fixed random lattice, it is in fact easier to solve the model (2), i.e. the annealed average over all random lattices, than to solve (1) on a regular lattice. This might be a reflection of the underlying reparametrization invariance of the theory at the critical point. One finds that the critical exponents of the models are changed relative to the values they have on a fixed regular lattice, and that this change agrees with calculations performed directly in a continuum framework, considering two-dimensional quantum gravity coupled to conformal field theories with central charges $c \leq 1$.

[1] One could also attempt to sum over all possible topologies, but we will not consider this possibility here.

[2] Of course one could also have placed the spins at the vertices in the triangulation. Historically they were placed at the triangles and we follow this convention.

It is still something of a mystery what happens when conformal field theories with $c > 1$ are coupled to quantum gravity. The continuum approach leads to complex critical exponents and it is presently not known how to repair this. Recently progress has been made in understanding the physics for $c > 1$ using the lattice approach [5, 6, 25, 27, 17]. Especially the limit $c \to \infty$ is now reasonably well understood. In addition it can be proven that the low temperature limit of the $q \to \infty$ Potts model agrees with the low temperature limit of the multiple Ising model when the number of copies $n$ of the Ising models goes to infinity [26]. Since the central charge of the multiple Ising system will be $n/2$ we see that the large $q$ limit of the Potts models provides information about the large $c$ limit.

At first sight this might be surprising since it is well known that the $q$-state Potts models has *first order* transitions for $q > 4$ on regular lattices while the multiple Ising models on regular lattices still have second order transitions. Second (or higher) order transitions allow us to think about divergent correlation lengths and associated continuum field theories. In the case of first order transitions we have no continuum field theory we can associate with the spin model and although a statistical model like (2) is perfectly well defined, we have *a priori* no continuum field theory for which we can say: (2) presents a regularized version of this field theory coupled to $2d$ quantum gravity. However, the study of critical properties of discretized $2d$ quantum gravity shows many examples of this kind. We can just mention the so-called multicritical matrix models. One observed that they gave rise to new critical behavior of the statistical system, but only later the corresponding conformal field theories which led to the critical exponents after coupling to gravity were identified as the $(2, 2m-1)$ minimal conformal field theories. From this point of view it seems justified to take the following attitude: if the annealed average of a $q$-state Potts model shows non-trivial critical behavior, i.e. exhibits a continuous phase transition rather than a first order transition, the corresponding critical behavior might indeed reflect the behavior of a conformal field theory coupled to gravity. The weak coupling results for $q \to \infty$ seem to support this idea and they indicate in addition that the behavior is associated with conformal field theories with $c > 1$ coupled to gravity for $q$ sufficiently large. This should serve as a strong motivation for a detailed study of the properties of the $q$-state Potts models coupled to $2d$ gravity. In addition the models have course some interest as statistical systems with new non-trivial critical behavior.

In this paper we will show that the critical behavior of the $q = 10$ states Potts model and the $q = 200$ states Potts model indeed changes from being trivial (i.e. first order) to non-trivial, namely second order for $q = 10$ and third order for $q = 200$, when we change from a regular lattice to the annealed average (2).

The rest of the article is organized as follows. Section 2 summarizes some recent predictions concerning large $q$ and $c$ models. Section 3 motivates the use of a particular class of triangulations $\mathcal{T}$. In section 4 we determine the order of the phase transition for the two Potts models coupled to gravity. In section 5 we discuss the back-reaction of the $q$-state Potts models on the geometry, defined by the piecewise linear manifolds corresponding to the triangulations, and we discuss the correlation between geometry and matter, more specifically between so-called baby universes and spin clusters. Finally section 6 summarizes our present understanding of the $q$-state Potts models coupled to $2d$ quantum gravity.

## 2 Expectations

Recently the first detailed predictions on the behavior of two-dimensional quantum gravity in the "strong coupling" regime have become available. In particular, the $\infty$-state Potts model and a class of $c = \infty$ models (including, for example, an infinite number of Ising species on the surface) have been solved exactly on a random surface using matrix model techniques. The biggest surprise was that all of these models are, up to some trivial transformations of the temperature, identical—and this inspite of the well-known fact that on a *fixed* two-dimensional lattice the Potts model has first-order, rather than continuous, phase transitions for $q > 4$.[3] The reason for this unexpected equivalence is that, roughly speaking, the $q = c = \infty$ model is a mean field theory: each state, each point in the (discrete) target space (the complete graph on $q$ points for the Potts model, an $n$-dimensional hypercube for $n$ Ising species) has so many neighbors that it feels only them, rather than the global structure of the target space.

The qualitative picture at $q = c = \infty$ is as follows. At high $\beta$ there is the usual ordered, pure gravity phase with $\gamma_s = -1/2$. In the low $q, c$ models this phase is bordered by the critical point (third-order) where $\gamma_s$ jumps to some higher value, followed by the low-$\beta$ phase—another phase, this time disordered, of pure gravity. In the case of $q = c = \infty$, however, at low $\beta$ the surfaces degenerate into trees, the nodes of which are spin clusters; in this phase the clusters are small and the trees are large, and therefore we have what is commonly called branched polymers, with $\gamma_s = +1/2$. At the multicritical point between the two phases we have $\gamma_s = +1/3$, and the transition is third-order, with $\alpha = -1$.[4] Thus for the $\infty$-state Potts model the transition is softened from first to third order by the coupling to gravity, while for the $c = \infty$ models the order is promoted from second to third.

[3]This explains the relative neglect of the random-surface Potts model at high $q$.

A direct approach to the branched polymer regime, one that avoids use of matrix models and deals directly with the triangulated surfaces, is possible [5, 6, 17]. Starting with theoretical arguments for branched structure at large $c$, as well as computer observations [9], one can *postulate* such a structure as a kind of a "polymerization" of an ordinary random surface model by restricting the boundaries of spin clusters to have minimal length. In this way, by assuming the spin cluster geometry of the $q = c = \infty$ surfaces, one arrives at the same picture of their critical behavior as via the exact matrix model solution, with the same values of the exponents $\gamma_s$ and $\alpha$. Finally, a novel combinatorial method [19] proves that trees are dominant in the $c \to \infty$ limit, and finds $\gamma_s = +1/2$ in the low-$\beta$ regime.

What about the Potts model with finite $q$? We know that at small $q$, the low-$\beta$ regime should be a disordered, pure gravity phase rather than a branched polymer. One way to get there is to expand the $q = \infty$ matrix model in inverse powers of $q$ [27]. The first non-trivial correction gives a new structure: at the vertices of the trees are no longer spin clusters, but complex graphs of spin clusters with (arbitrarily many) loops, structures which we will call "galaxies." By adjusting the coupling constants one can now vary three things: the area of the spin clusters, the size of the galaxies (i.e., the number of clusters inside the galaxies), and the size of the trees (the number of galaxies inside the tree).[5]

In this approximation, for large enough $q$ the critical behavior is as follows: there is the ordered, pure gravity phase ($\gamma_s = -1/2$) at large $\beta$, a phase of large spin clusters[6]; there is still a third-order transition ($\alpha = -1$) at some $\beta_{c2}$ to a branched polymer phase of large trees that has $\gamma_s = +1/2$; but now, for any finite $q$ at $\beta = \beta_{c1}$ there is another third-order transition ($\alpha = -1$) to a phase where the galaxies are large and $\gamma_s = -1/2$ (once again), and which extends all the way down to $\beta = 0$; finally, at both transitions $\gamma_s = +1/3$. This new large galaxy phase can be identified as an disordered phase—precisely what is missing at $q = \infty$. As $q$ decreases from infinity, $\beta_{c1}$ and $\beta_{c2}$ approach each other: the branched polymer, mean field phase gets eaten away, precisely as expected; until some $q_c$ at which the branched polymer phase disappears altogether, leaving us with just the picture we know to be true at low $q$: two pure gravity phases, an ordered and a disordered one, separated by a phase transition. Fig. 1 is a schematic two-dimensional slice of this phase diagram.[7]

[4] It should be noted that here, as well as in the Ising model, there are no logarithms in the free energy and—as $\alpha$ is an integer—no poles in any of its derivatives: there is simply a discontinuity in the third and all higher derivatives at the critical point.

[5] Subsequent refinements to this approximation should not add any new essential features.

[6] A "large" quantity in this context means that some of its moments diverge.

[7] The third dimension that has been suppressed is the cosmological constant: in simulations of

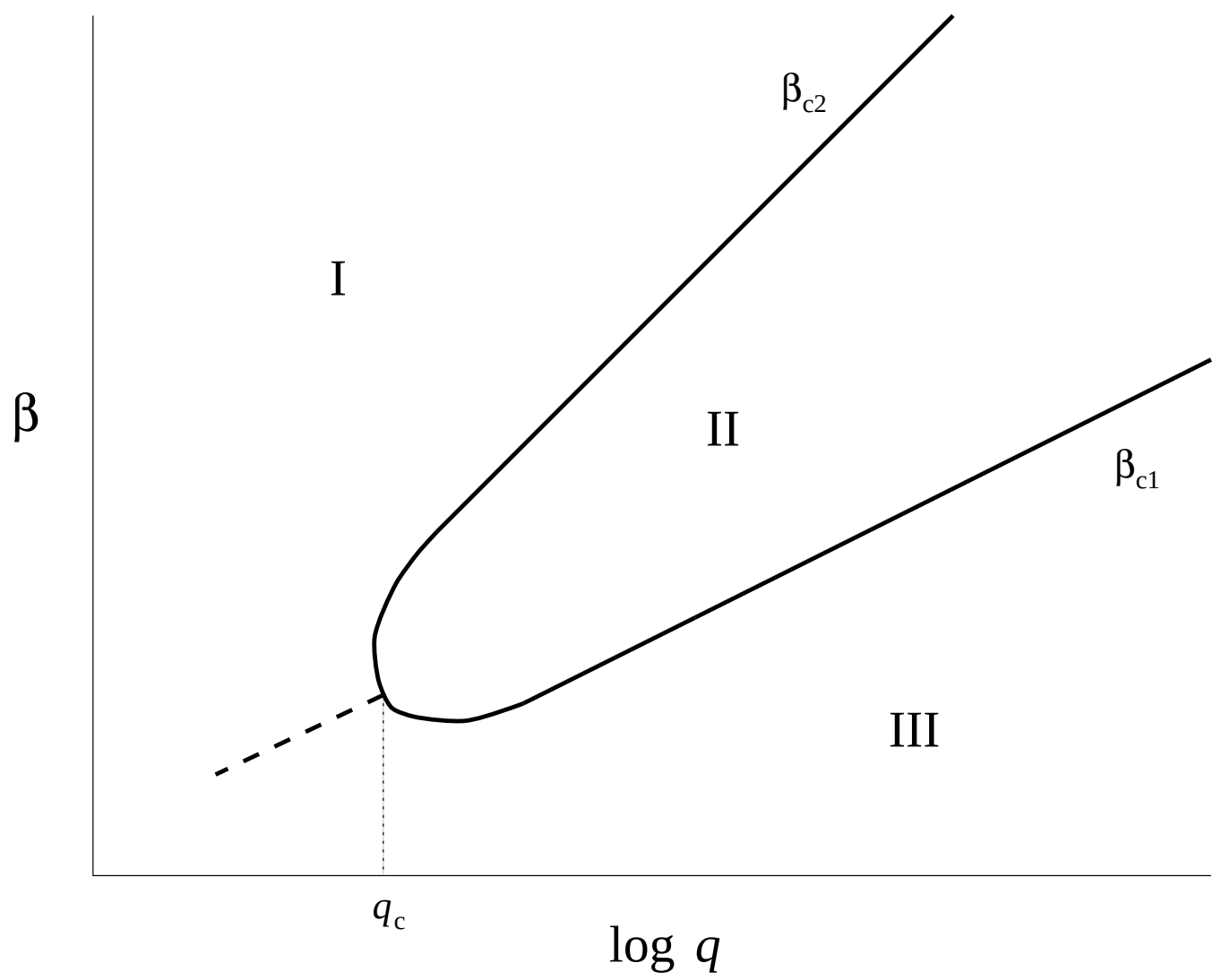

Figure 1: Schematic phase diagram of possible phase diagram of the Potts model at large $q$. (I) is the ordered (pure gravity) region of large clusters, (II) is the branched polymer region of large trees, and (III) is the disordered (pure gravity) region of large galaxies.

The exact critical values of the parameters of course depend on the truncation used in the above approximation (see [27] for details). Two methods were used in [27] that gave $q_c \approx 120 \pm 20$. The same two methods give $\beta_{c2} - \beta_{c1} \approx 0.1 \pm 0.05$. The asymptotic behavior of $\beta_{c2}$ is $\frac{1}{2}\log\Delta + \mathcal{O}(\Delta^{-1})$, where $\Delta = q - 1$. $\beta_{c1}$ is harder to calculate, but it seems that for $\Delta$ large enough $\beta_{c1} \approx \frac{1}{4}\log\Delta$.

## 3 The use of tadpoles

It is by now well established that Monte Carlo simulations may serve as an effective tool for extracting the critical behavior of systems like (2). Let us here only mention that the Monte Carlo simulations involve two types of updating: we have to update the fluctuating random triangulation and we have, for a given random lattice, to update the matter fields. The updating of the triangulation uses the so-called flip, where two adjacent triangles, viewed as a square with the common link as the diagonal, is replaced by the same square but made out of the two triangles formed by using the other diagonal in the square (the diagonal is "flipped"). For a given triangulation the updating of the spins is performed by the Swendsen-Wang cluster algorithm. In most simulations used until now the class of allowed triangulations has been characterized as combinatorial two-manifolds. The two requirements satisfied large surfaces, this can of course be considered tuned to its critical value.

by such triangulations are that any two vertices have at most one link in common and that all vertices, which are not on the boundary, belong to at least three links. It can be convenient to relax this concept of a combinatorial manifold and consider all triangulations which can be obtained by successive gluing of triangles along their sides to form a closed surface of spherical topology, i.e. so that $N_T - N_L + N_V = \chi = 2$, where $N_T$, $N_L$ and $N_V$ denote the number of triangles, links and vertices in the closed surface after the gluing. For such triangulations we can have one-loops (i.e. links which start and end at the same vertex) and two-loops (i.e. two vertices joined by more than one link as mentioned above). To each triangulation we can associate a dual $\phi^3$ graph; the difference between the $\phi^3$ graphs which correspond to combinatorial two-manifolds and the $\phi^3$ graphs which correspond to the spherical triangulations obtained by the unrestricted gluing is that the last set of graphs contains tadpole and self-energy diagrams which are associated with the one- and two-loops.

Whether or not one- and two-loops are included should not effect the scaling limit of the theory, and we know that it does not in the cases where the theories can be solved explicitly, i.e. for central charge $c < 1$. The non-universal features are affected, though. Especially the magnitude of finite-size effects can be different. This is important from a practical point of view and we have observed this difference both in strong coupling expansions and Monte Carlo simulations. In both cases the inclusion of tadpoles and self-energy diagrams significantly reduces the finite-size effects.

Let us first discuss the strong coupling series. We can expand the free energy up to order 14 in the cosmological constant $g$, and use a [2,1] Padé approximant on the ratios of successive coefficients, which should behave as $g_c^{-2}(n/n-2)^{\gamma_s - 3}$. For pure gravity, where $\gamma_s = -1/2$ for spherical graphs, we find $\gamma_s = -0.503$ when we include all graphs, but $\gamma_s = -0.516$ when we exclude tadpoles and self-energy diagrams. The relative advantage of tadpoles and self-energy diagrams becomes more marked if, say, an Ising model (i.e. a $q = 2$ Potts model) is coupled to the surface. In fig.2 we plot $\gamma_s$ as a function of $c = e^{-2\beta}$ using the Padé approximation: the benefit of tadpoles and self-energies is evident. Other Padé approximants and other extrapolations like the ratio method give similar results.

In order to test this effect directly on much larger surfaces we have performed Monte Carlo simulations with high statistics ($10^6$ sweeps) on the Ising model at the critical values $\beta_c$ for both classes of triangulations . The results are shown in fig. 3. Finite-size scaling is used to extract the critical indices related to the magnetisation $M$ and the susceptibility $\chi$. The comparison is made easier by the fact that the exact critical coupling for the Ising model is known for both types of triangulations, $\beta_c = \frac{1}{2}\log(108/23)$ if one and two-loops are excluded and $\beta_c = \frac{1}{2}\log((2\sqrt{2}-1)/27)$ if they are allowed [15, 16].

The simulations are performed on triangulations ranging from $N = 20$ to $N = 20480$, and from the magnetisation $M$ and the susceptibility $\chi$ finite size scaling allows us to extract the combinations $\beta/\nu d_H$ and $\gamma/\nu d_H$ of the critical exponents. $\beta$ and $\chi$ are the critical exponents of the spin and the susceptibility while $\nu$ is the exponent for the correlation length. $d_H$ denotes the Hausdorff dimension of the ensemble of piecewise linear surfaces defined by the triangulations.

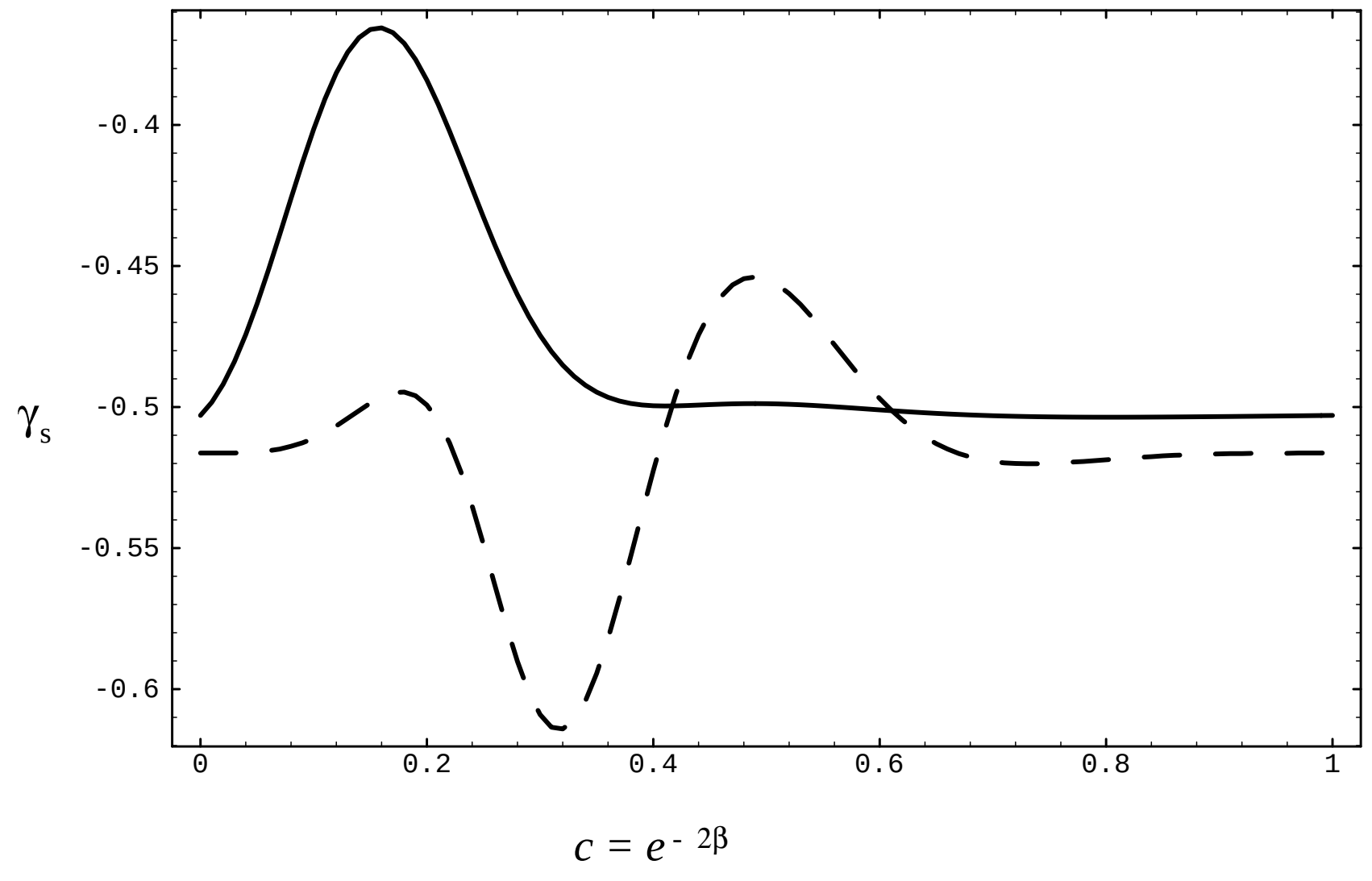

Figure 2: $\gamma_s$ as a function of temperature for the Ising model, using strong-coupling expansion. Solid curve is with tadpoles and self-energies included, dashed curve is with them excluded.

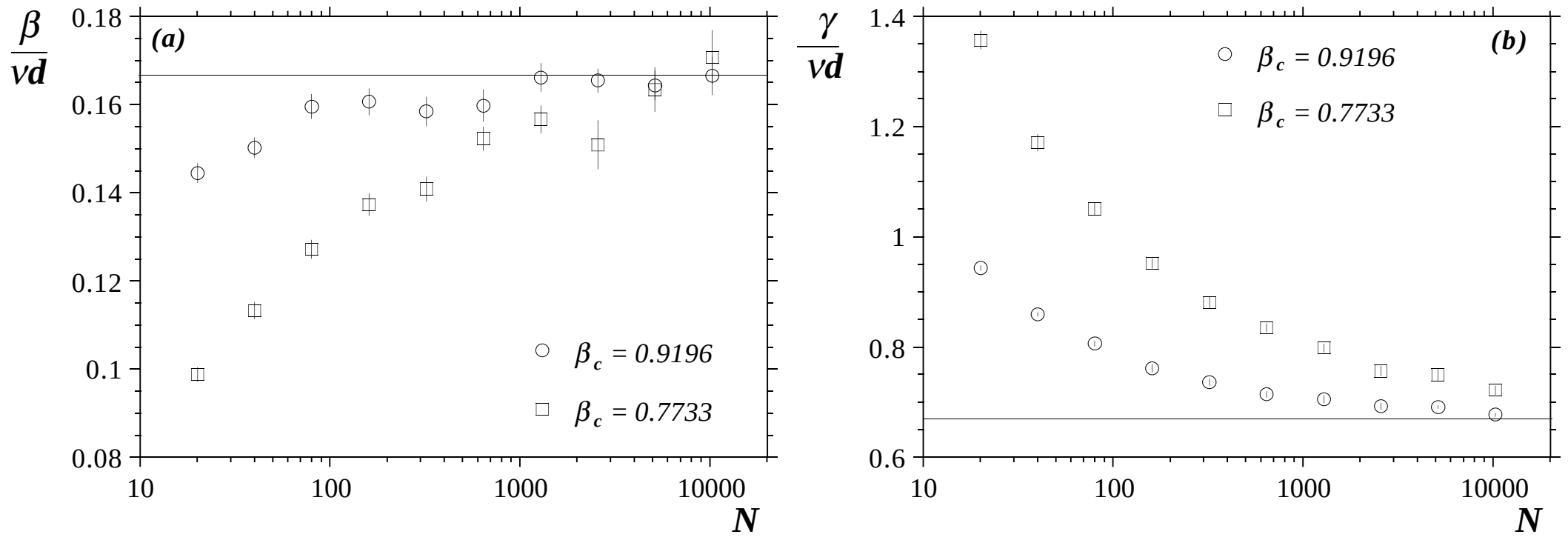

Figure 3: The extracted values of (a) $\beta/\nu d_H$ and (b) $\gamma/\nu d_H$ (measured at $\beta_c$) versus lattice size for one Ising model coupled to gravity. The data points with circles are obtained for triangulations which include one- and two-loops and the data points with squares are obtained for triangulations where the one- and two-loops are excluded. The solid lines represent the analytical values $\beta/\nu d_H = 1/6$ and $\gamma/\nu d_H = 2/3$.

From the finite size scaling relations it follows that $\beta/\nu d_H$ and $\gamma/\nu d_H$ in principle can be determined from two consecutive lattice sizes $N$ and $2N$, i.e.

$$\frac{\beta}{\nu d_H}(N) = \frac{\log M(2N) - \log M(N)}{\log 2}, \qquad \frac{\gamma}{\nu d_H}(N) = \frac{\log \chi(2N) - \log \chi(N)}{\log 2}. \tag{3}$$

In fig. 3 we plot the extracted values versus lattice size. It clearly demonstrates how the finite size corrections are diminished if loops of length one and two are included.

Another motivation for using the triangulations with one- and two-loops comes from the conjecture [7] that the transition to the magnetized phase takes place only after the surfaces have developed a pronounced tree-like structure where the individual branches are already magnetized *before* the transition, while the total magnetisation is still zero. The transition to the magnetized phase is only the final alignment of the spins of the individual baby universes. Allowing one- and two-loops makes it easier to detect how the location of spin clusters fits to the baby universe distribution as the fractal structure becomes more pronounced.

## 4 Order of the phase transitions for $q = 10$ and $200$

### 4.1 The simulations

We now turn our attention to the simulations of the $q$-state Potts models for $q = 10$ and $q = 200$. As for the Ising model we use the flip algorithm to update the triangulations and the Swendsen-Wang cluster algorithm for the spins. The lattice sizes were $N = 250$, 500, 1000, 2000, 4000 and 8000. The main problem in the simulations is that the autocorrelations seem to grow linearly with $q$, presumably due to the increased branching of the surfaces.[8] This forced us to perform very long runs in order to get good statistics. In the vicinity of the phase transition we performed approximately $5 \times 10^6$ and $2 \times 10^7$ sweeps for each $\beta$ value for the $q = 10$ and $q = 200$ state Potts model, respectively. With the autocorrelations present this resulted in between 150.000 and 400.000 independent measurements for $q = 10$, and 40.000 to 60.000 independent measurements for $q = 200$, depending on lattice size.

### 4.2 Cumulants

It can be difficult to distinguish numerically between a continuous and a first order phase transition. Many methods have been proposed [14] which either depend on the presence of hysteresis at $1st$ order transitions or use finite size scaling. All of the methods have their drawbacks. Hysteresis can be hard to detect in simulations and if one drives the system too fast pseudo-hysteresis can also be seen in second order transitions and one quickly enters into discussions of weak $1st$ order transitions versus strong $2nd$ order transition. Finite size scaling only applies when the system size is larger than the characteristic correlation length at the pseudo-critical transition point and is based on a number of assumptions which are not necessarily satisfied. Especially in the case where we consider matter systems coupled to quantum gravity one should consider the assumptions with some suspicion: what is the system size when we integrate over fluctuating geometries of fixed volume? Recall the basic equation of finite size scaling:

$$L^d f_L(t) = F(t^\nu L), \qquad t = |\beta - \beta_c| \tag{4}$$

where $L$ is the size of a regular lattice, $f$ the bulk free energy density, $\beta$ the inverse temperature, $d$ the dimension of the the lattice and $\nu$ the critical exponent of the correlation length $\xi(\beta) \sim 1/|\beta - \beta_c|^\nu$ In our case we have only $N$, the total volume. We should presumably make an identification $N = L^{d_H}$ where $d_H$ is the Hausdorff dimension of our ensemble of manifolds, and we should *presumably* view $\xi(\beta)$ as some average *geodesic* distance on the same ensemble, but it is clear that the foundation for finite size scaling is not as firm as on regular lattices. We have nevertheless chosen to use finite size scaling for two reasons: as was clear from the data presented in the last section it seems to work very well for matter fields with central charge $c < 1$ coupled to gravity so it is not a vain hope that it will work well for other models. Secondly it is the method which allows us to extract critical exponents in the cleanest way.

[8] It would be very interesting to use the new algorithms developed [21] to deal with surfaces of pronounced fractal structure on this problem. However, the algorithms were not available when this project was started.

Let us now look at scaling of the specific heat $C_V$ and the energy cumulant $V_L$

$$V_L = 1 - \frac{\langle E^4 \rangle}{\langle E^2 \rangle^2}. \tag{5}$$

In the case of a first order phase transition the specific heat will diverge and in finite systems its maximum will grow linearly with lattice size, $C_V(max) \sim N$. In fig. $4a$ and $b$ are shown the measurements of the specific heat both for $q = 10$ and $q = 200$. The maximum increases very slowly with lattice size, much slower than is expected for a $1st$ order transition. These statements are quantified in fig. $5a$ where $C_V(max)/N$ is plotted versus $N^{-1}$ for both models. If we were dealing with first order transitions this quantity should approach a non-zero constant, but instead it seems to go to zero.

The energy cumulant $V_L$ provides even more conclusive evidence. For a continuous phase transition it vanishes in the infinite volume limit for all values of $\beta$, while at a $1st$ order transition fluctuations between coexisting phases prevent it from becoming zero in the critical point [12, 13, 18, 20]. In that case $V_L$ reaches a non-zero minimum value which scales with $N$. In fig. $4c$ and $d$ we show $V_L$ for different lattice sizes for $q = 10$ and 200. In both cases it approaches zero as $N$ increases and no evidence of the characteristic behavior of a $1st$ order phase transition (a persistent dip) is seen.

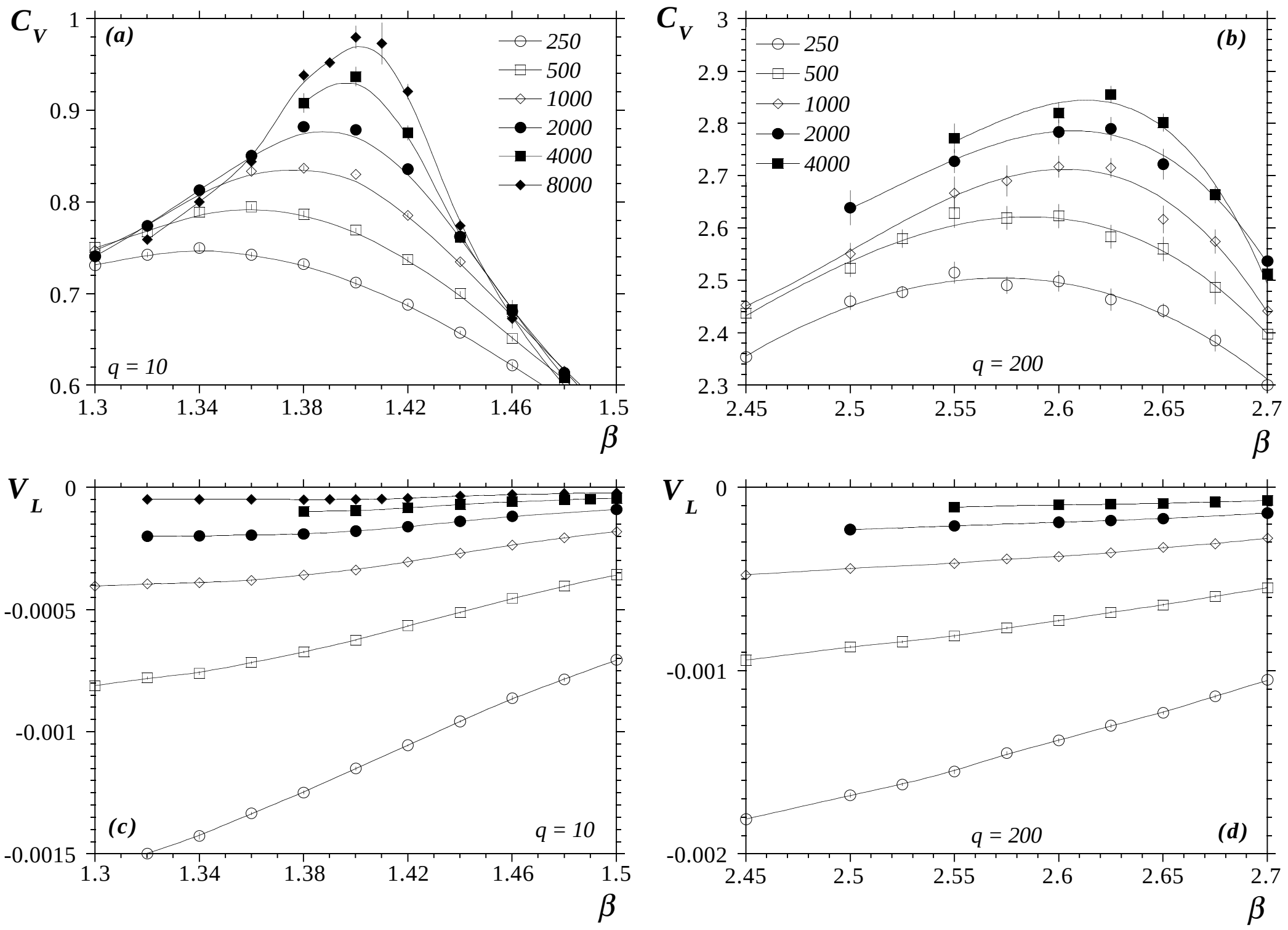

Figure 4: The specific heat $C_V$, for various lattice sizes, for (a) $q = 10$ and (b) $q = 200$ state Potts models coupled to gravity. In (c) and (d) the energy cumulant $V_L$ for the same models is shown.

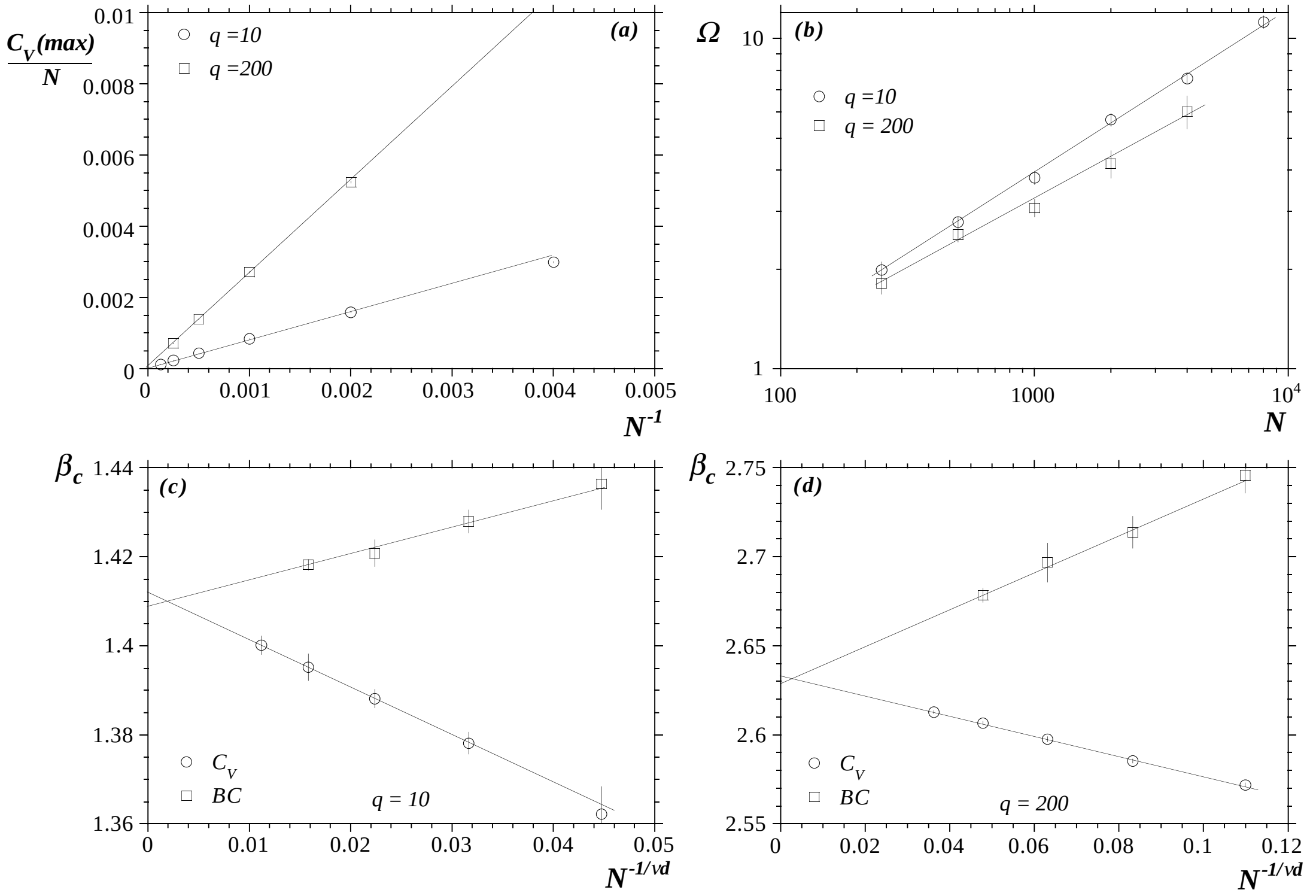

Figure 5: (a) $C_V(max)/N$ versus $1/N$ for $q = 10$ and 200. Linear fits show that $C_V(max)/N \to 0$ as $N \to \infty$. (b) The maximal slope of Binder's cumulant versus lattice size. The linear fits yield $\nu d_H$. (c) and (d): the pseudo critical points $\beta_c(N)$ versus $N^{-1/\nu d_H}$, both obtained from $C_V(max)$ and from intersection of BC.

Taken together these results provide strong evidence in favor of *continuous* transitions. As already emphasized there are no compelling arguments why a $1st$ order phase transition on a regular lattice stays first order after we perform the annealed average (2) over random triangulations. On the contrary, for the Ising model it is known that the phase transition is weakened from $2nd$ to $3rd$ order, and numerical simulations show that the same is the case for the 3-state Potts model. It is true that in earlier simulations on the 10-state Potts model coupled to gravity it was claimed that the transition was $1st$ order [11]. But that was concluded from the qualitative behavior of Binder's cumulant, which is not a very good indicator to the order of phase transitions. In fact our measurements are in excellent agreement with those in [11], but we must draw a different conclusion.

## 4.3 The critical temperature

The next step is to locate the critical temperature. It can be done in two ways by finite size scaling. One can either use the location of the peak in the specific heat $\tilde{\beta}_c$, which scales as

$$| \tilde{\beta}_c - \beta_c | \sim N^{-1/\nu d_H}, \qquad (6)$$

or look at the intersection of Binder's cumulant for different lattice sizes.

Let us first consider the specific specific heat. As we do not have many lattice sizes at our disposal it would be impossible to make a three-parameter fit to (6). Fortunately it is possible to determine $\nu d_H$ directly, as the maximum of slope of Binder's cumulant scales as

$$\max\left(\frac{\partial BC}{\partial \beta}\right) \sim N^{\nu d_H}, \qquad (7)$$

where Binder's cumulant is defined as

$$BC = \frac{\langle M^4 \rangle}{\langle M^2 \rangle^2} - 3 \qquad (8)$$

In fig. 5*b* we show the lhs of (7) on a log-log plot and the corresponding linear fits. The values of $\nu d_H$ obtained in this way are listed in table 1. Knowing $\nu d_H$ we can now make a linear fit to (6) and obtain $\beta_c$ ($\beta_c(C_V)$ in table 1).

Next we turn to the use of Binder's cumulant (8) to locate the critical point of the two multiple Potts models. As is well known the curves of $BC$ intersect at $\beta_c$ up to finite size corrections. Our statistics are so good that we can actually see how the intersection point moves with increasing lattice size. It is to be expected that the intersection point moves towards the true critical point in the same way as the maximum in the specific heat, i.e. as given by (6). This is shown, together with $\beta_c(C_V(max))$, in fig. 5*c* and *d*. Extrapolating to $N = \infty$ yields the critical points shown in table 1. The two methods for determining $\beta_c$ are in fairly good agreement, the slight discrepancy in the values is presumably being due to the fact that larger lattices are needed before the methods converge completely.

It should finally be mentioned that the critical temperature we find agrees fairly well with the asymptotic prediction [25] $\beta_{c_2} = \frac{1}{2}\log q + O(q^{-1}) \asymp 2.65$.

## 4.4 Critical exponents

To determine the magnetic critical exponents $\beta$, $\gamma$ and $\alpha$ we apply finite size scaling at the critical point (taking $\beta_c$ as the weighted average of $\beta_c(C_V)$ and $\beta_c(BC)$). From the scaling of the magnetization $M$ and its derivative with respect to $\beta$ (which scales as $N^{(1-\beta)/\nu d_H}$) we get estimates for $\beta$ and $\nu d_H$. ¿From the scaling of the magnetic susceptibility we get $\gamma/\nu d_H$. Finally, and most interesting, the specific heat allows us to estimate $\alpha$, both from the scaling of its maximal value and at the critical point. All these exponents are shown in table 1. The fits to the specific heat are shown in fig. 6. For $q = 10$ $\alpha$ is very close to zero and we have thus tested whether the specific heat fits better to a logarithmic divergence

$$C_V(N) = c_1 + c_2 \log(N). \tag{9}$$

| $q$ | $\nu d^{(a)}$ | $\nu d^{(b)}$ | $\beta_c(C_V)$ | $\beta_c(BC)$ | $\beta$ | $\gamma$ | $\alpha^{(a)}$ | $\alpha^{(b)}$ |
|---|---|---|---|---|---|---|---|---|
| 10 | 2.01(5) | 2.03(4) | 1.412(1) | 1.409(2) | 0.53(1) | 1.11(2) | +0.02(4) | +0.02(3) |
| 200 | 2.49(15) | 2.52(12) | 2.633(1) | 2.629(3) | 1.18(4) | 0.31(5) | −0.83(6) | −0.80(5) |

Table 1: The extracted critical points and exponents for the $q = 10$ and 200 state Potts models coupled to gravity. The values are obtained using finite size scaling. For $\alpha$ we have values both from the scaling of (a) $C_V(max)$ and (b) $C_V(\beta_c)$, and for $\nu d_H$ from the scaling of (a) $\max\{\partial BC/\partial\beta\}$ and (b) $M$ and $\partial M/\partial\beta$. The main contribution to the errors comes from the location of $\beta_c$.

This fit is also shown in fig. 6 and turns out to be equally good as the one with $\alpha = 0$.

### 4.5 Comments

What is the interpretation of these results? The first thing to remember is that for the $q = 4$ state Potts model coupled to gravity we have $\beta = 1/2$, $\gamma = 1$, $\alpha = 0$ and $\nu d_H = 2$, i.e. consistent with our measurements for the 10-state Potts model. It is natural to conjecture from this coincidence that we have an interval of $q$'s starting at $q = 4$ and extending beyond $q = 10$ where we have *identical* critical behavior after coupling to gravity. We will discuss this remarkable possibility later.

At some large value of $q$ the the phase transition changes character. $\alpha$ has a value close to $-1$ for $q = 200$. This value agrees both with [27] and [6], indicating that we might have reached the large $q$ limit in which the magnetic transition takes place between surfaces degenerated to branched polymers (in the phase with no magnetization) and pure gravity (the low temperature phase where the system is magnetized). We will later show that the entropy exponent for the surfaces gives some support to this picture.

Whether this change of critical behavior happens at some critical value of $q$, or whether there is a gradual change from one behavior to another cannot be established from our data.

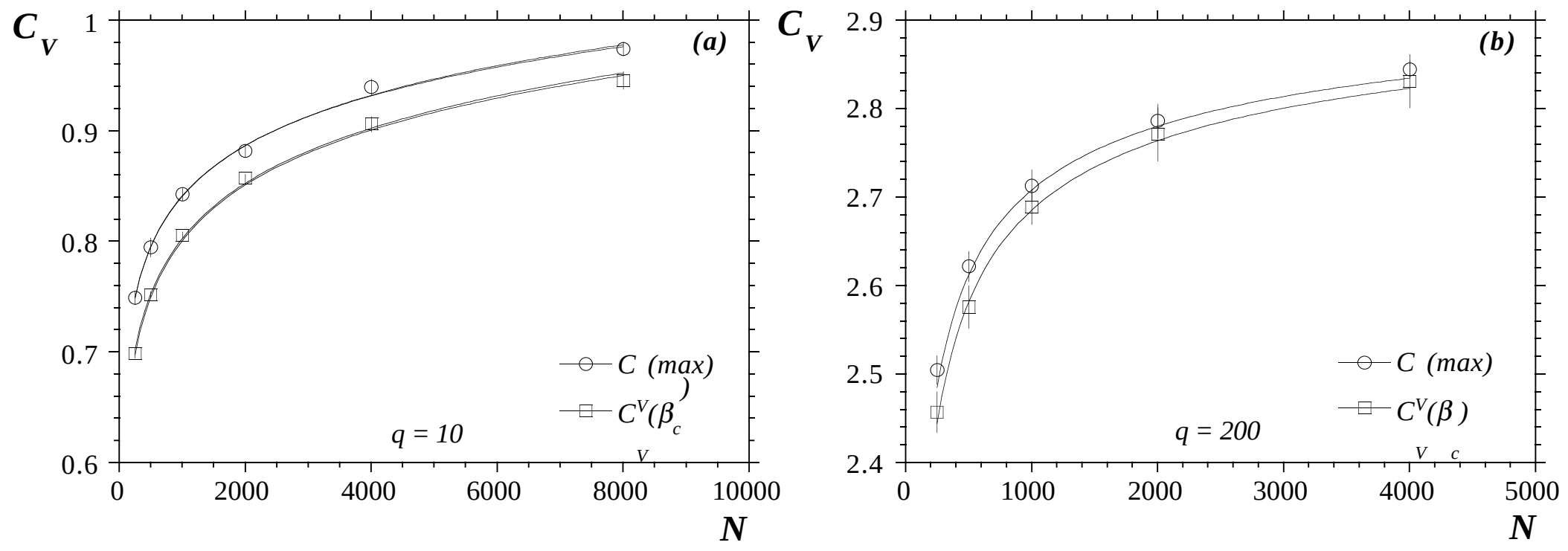

Figure 6: The measured values of $C_V$, both the maximal value and in the critical point $\beta_c$, for (a) $q = 10$ and (b) $q = 200$ state Potts model. Included are fits to $C_V = c_0 + c_1 N^{\alpha/\nu d_H}$ and, for $q = 10$, an equally good fit to $C_V = c_0 + c_1 \log N$

# 5 The fractal structure of surfaces and spin-clusters

## 5.1 Expectations

Let the partition function of the annealed average of the Potts models over random triangulations be given by (2). If we perform the summation over spin configurations we can write the partition function as

$$Z(\beta, N) \sim N^{\gamma_s(\beta)-3} e^{\mu_c(\beta) N} \qquad N \to \infty. \tag{10}$$

This formula defines the string susceptibility exponent $\gamma_s(\beta)$. It is tacitly assumed that the partition function *can* be written in this form and it is well known that it is the case for $q = 2$ and $q = 3$ where the models can be solved analytically.

Recall the scenario of the critical behavior of the large $q$ Potts model presented in sec.2. We have tried to substantiate this picture in the case of $q = 200$. There are three questions to address:

1. Can we observe the branched polymer phase for some temperature range just above the transition from a magnetized phase to a phase with total magnetization equal zero?
2. Can we observe a second phase transition at higher temperature from the branched polymer phase and back to some pure gravity phase?
3. Can we observe the magnetisation of baby universes and the correlation between spin clusters and baby universes?

## 5.2 Search for branched polymers

Let us first look for the branched polymer phase. The exponent $\gamma_s$ is a good signal. It should be equal 1/2 if the surfaces are degenerate into genuine branched poly-mers. Another quantity, suggested in [7] is the maximal "link" radius $\langle r_{max} \rangle$. More precisely we define a distance $d_{ij}$ between two vertices $i$ and $j$ in a triangulation as the length of the shortest path along links between the vertices. $r_{max}(T)$ is now defined by

$$r_{max}(T) = \frac{\sum_i^{V_T} \max_j \{d_{ij}\}}{V_T}, \tag{11}$$

where $V_T$ denotes the number of vertices in the triangulation $T$. $\langle r_{max} \rangle$ is now the average value of $r_{max}(T)$ in the ensemble of triangulations defined by the partition function $Z(N, \beta)$ from (2). Although $\langle r_{max} \rangle$ is not a universal quantity like the susceptibility exponent $\gamma_s$ it reflects very well the geometry of the surfaces and it is known that in the case of multiple $q = 2, 3$ and 4 state Potts models coupled to gravity it shows a pronounced dip on the high temperature side of the spin transition, the depth increasing with the total central charge of the system.

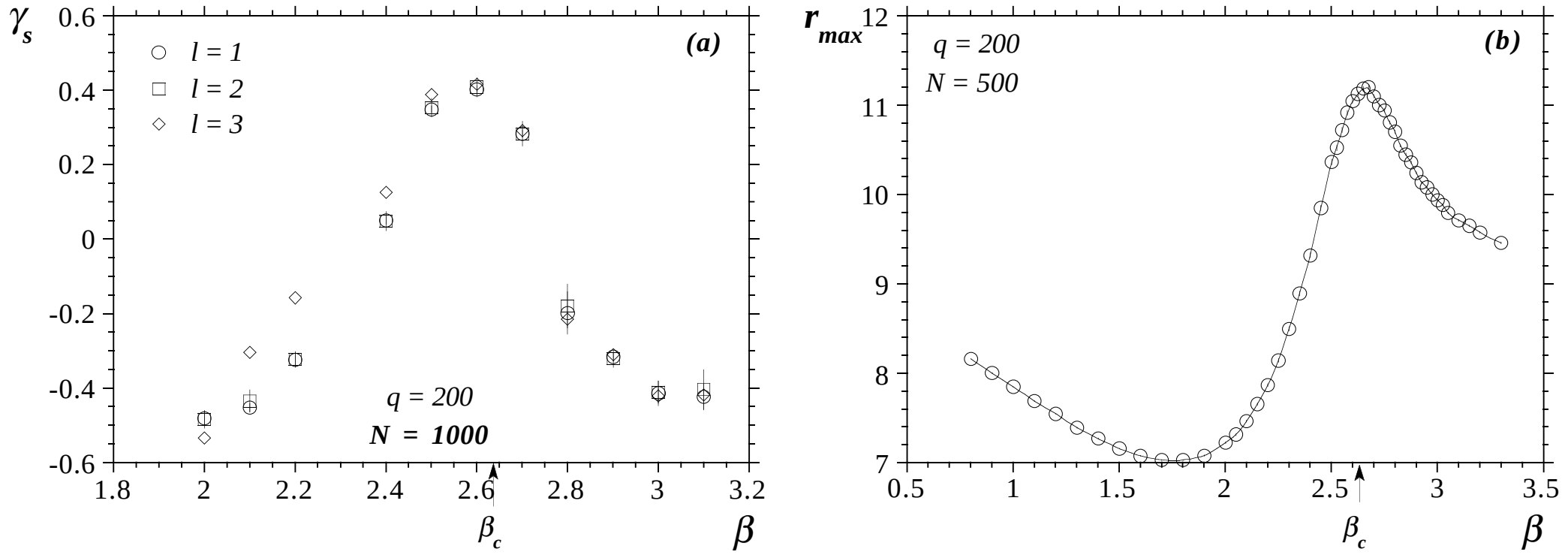

Figure 7: Measurements of $\gamma_s$ and $r_{max}$ for $q = 200$ state Potts model coupled to gravity.

In fig. 7 we plot these quantities for $q = 200$ and a wide range of $\beta$. The values of $\gamma_s$ are determined from the counting of the so-called baby universes. For an ensemble of surfaces given by the partition function (10) the distribution of baby universes with area $B \leq N/2$ will be given by[9]

$$n(B) \sim N B^{\gamma_s - 2} \left(1 - B/N)^{\gamma_s - 2}\right). \tag{12}$$

$\gamma_s$ is now determined by fitting the measured distribution to (12). The measurements of $\gamma_s$ may indicate a branched polymer phase starting at $\beta \approx 2.4$ and extending to the magnetisation transition. We have a rather broad peak value around $\gamma_s \approx 0.4$. This broad peak is in sharp contrast to the very narrow peaks one observes for spin systems which at the critical point corresponds to a conformal field theory with $c < 1$ (i.e. here $q = 2$ and 3) and it gives some indication that there might be a plateau where $\gamma_s \neq -1/2$. (For $q = 2$ and 3 it is known that $\gamma_s$ is only different from the pure gravity value $-1/2$ precisely *at* the transition point). We do not get the precise value $\gamma_s = 1/2$ for branched polymers. It is not clear to us whether this is finite size effects or simply the fact that we are below the critical value $q_c$ above which we should enter the branched polymer phase. There is the following theoretical reason for believing that we observe finite size effects: from general arguments spin systems should either have $\gamma_s \leq 0$ or $\gamma_s = 1/(n+1)$ [17], and probably $N = 1000$ is not very large for $q = 200$. Unfortunately it is very time consuming to enlarge the system for $q = 200$, as already mentioned.

[9]As we allow triangulations with one-loops and two-loops we can just as well extract $\gamma_s$ from the distributions of baby universes with neck length one and two as well as length three (as was used in the original measurements [8]). We performed simulations on pure gravity in order to test this. The results for surfaces of size $N = 1000$ were: $\gamma_s(l = 1) = -0.501(6)$, $\gamma_s(l = 2) = -0.502(8)$, $\gamma_s(l = 3) = -0.498(22)$. We conclude that there is universality for $n(B, l)$ as long as the loop length $l$ is very small.

When we look at $\langle r_{max} \rangle$ it is important to keep in mind that it is not a universal quantity. It might change quite a lot without reflecting a corresponding change in the critical properties of the system. In fact it decreases continuously from the high temperature limit, where it agrees with the value for pure gravity to a minimum at $\beta \approx 1.75$. After this it increases rapidly to its maximum close to $\beta_c$, corresponding to the magnetisation transition. It does not fit in a very convincing way with the behavior of $\gamma_s$, but one *could* argue that it is not inconsistent with the following picture: at infinite temperature the spin system has a life independent of the underlying lattices, which therefore have the geometry of pure gravity. As the temperature decreases the interaction between lattices and the spin system increases and the lattices are deformed towards branched polymers[10]. Only deep into this phase $\gamma_s$ will then jump from $-1/2$ to $1/2$. This suggests that one should look for the second transition suggested in [27] in the region $1.75 < \beta 2.5$.

### 5.3 Search for the second transition

According to the conjecture in [27] the second transition, where $\gamma_s$ jumps back from $1/2$ to $-1/2$, while taking the value $-1/3$ at the transition point should be a $3rd$ order transition. From the point of view of the fractal geometry of the surfaces it is a kind of mirror transition of the magnetisation transition, but taking place at somewhat higher temperature. Clearly it cannot be a spin transition in the usual sense, so the only general signature we have is that the critical exponent of the specific heat $\alpha$ should be $-1$, i.e. there should be a discontinuity in the derivative of the specific heat. We have scanned values of $\beta$, especially the region suggested above, but have found no trace of a transition.

[10]More precisely the dual lattice will acquire a conventional branched polymer structure since the spins are placed on the dual lattice. This is precisely what gives a decreasing value of $\langle r_{max} \rangle$, for the triangulated surfaces [7].

## 5.4 Spin clusters

Let us now turn to the correspondence between spin clusters and the fractal structure. From [25, 6, 27] the magnetisation transition for large $q$ or large $c$ arises by alignment of already magnetized baby universes. To which extent is that true? To answer that question we propose two new observables:

- The ratio between number of boundary links $b_l$ (links separating different spin clusters) and number of spin clusters $N_C$:

$$W = \frac{b_l + 1}{N_C}. \tag{13}$$

- The number _3 of sets of three clusters that all share boundaries with each other which we shall call triplets (normalized by the number of clusters).

Let us explain why these observables tell us something about the alignment of spin clusters. First $W$. It is always larger or equal than one and, more important, it can only be equal to one in two situations: $(i)$ if the system is magnetized and hence there is only one cluster (no boundary links), $(ii)$ if all the clusters are glued together in a treelike structure separated by boundaries of length one. Similarly for $\lambda_3$: It can only be zero in the two situations described above. It is however not as sensitive to the neck size separating the baby universes. So if the baby universes are completely magnetized before the phase transition we should observe $W \approx 1$ and $\lambda_3 \approx 0$.

Fig. 8 shows the measurement of these quantities for $q = 4$, 10, 50 and 200 state Potts models coupled to gravity and, for comparison, for the same models on a regular square lattice.

Let us start with a discussion of $W$ (fig. 8 $(a)$). On a triangulated surface each spin has three neighbors so that for very high temperatures, where the spins fluctuate independently, each spin will form its own cluster (for $q$ large enough) and every link is a boundary link. The ratio between links and spins on a triangulation is 3/2, which is what we measure in the $\beta \to 0$ limit. For $\beta \gg \beta_c$ there is only one cluster, hence $W = 1$. What is interesting is that the value around the magnetization transition decreases as $q$ increases. It starts at $W \approx 1.5$ for $q = 4$ and reaches $W \approx 1.05$ for $q = 200$. These values do not change notably with lattice size. To distinguish between situations $(i)$ and $(ii)$ we measure the number of spin clusters around $\beta_c$. For $q = 200$ $N_c \approx N/4$, *which shows that* $W \approx 1$ *because the baby universes are magnetized individually (in different direction) while the total magnetization is still zero.* In other words, the graph of spin clusters is large and increasingly tree-like at the critical point as $q \to \infty$.

Very different behavior is seen on a regular lattice. As before the spins fluctuate independently at high temperature and we have $W \approx 2$ (the ratio between links and spins on a square lattice). For high $\beta$ we get $W = 1$ due to total magnetisation.

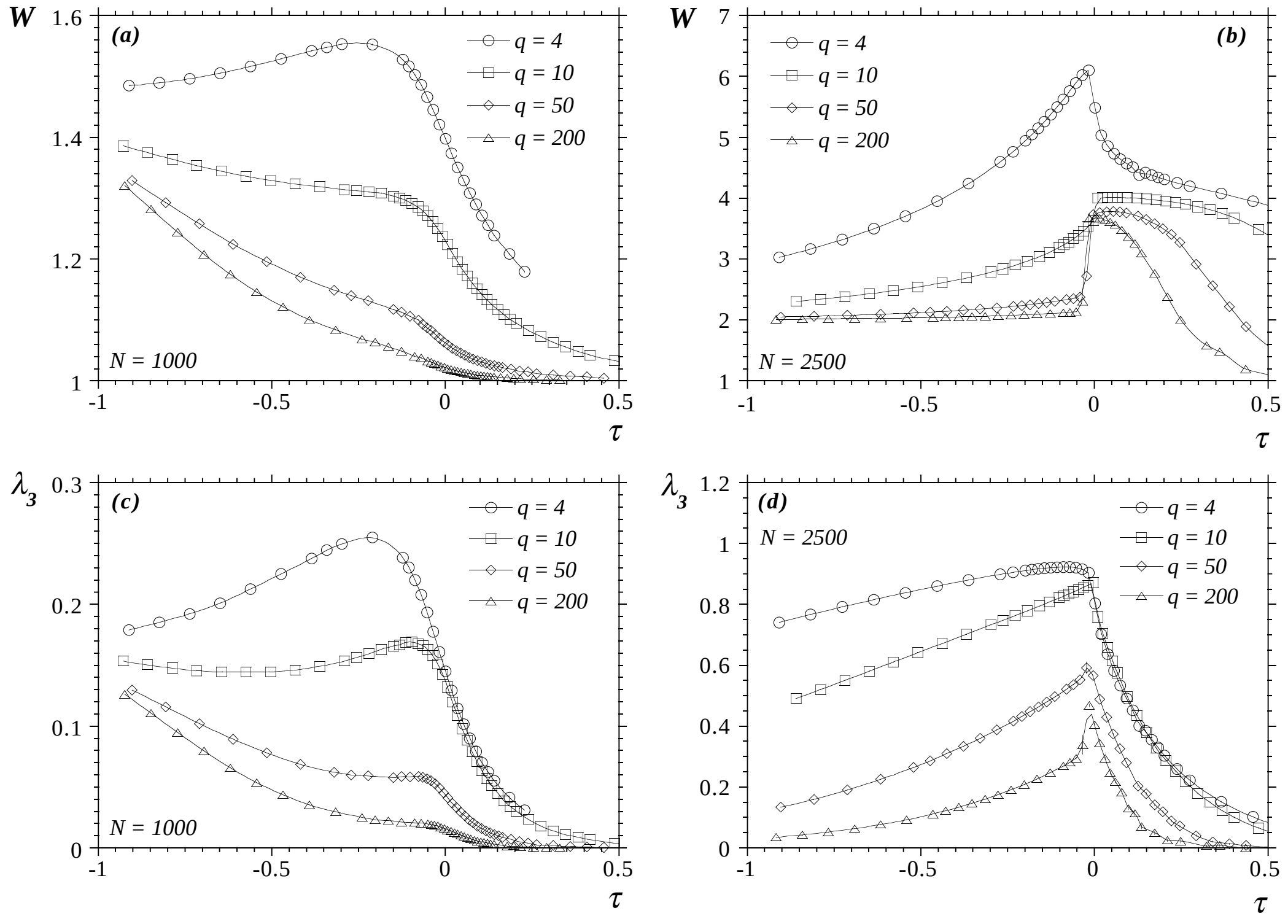

Figure 8: Measurements of $W$ and $\lambda_3$ for $q = 4$, 10, 50 and 200 state Potts models coupled to gravity ((a) and (c)) and, for comparison, on a regular lattice ((b) and (d)). The values are plotted versus the reduced temperature $\tau = (\beta - \beta_c)/\beta_c$.

But close to the phase transition the value of $W$ is very large as there are large and complicated clusters (and which cannot decrease the boundaries by deforming the lattice). Furthermore the sharp jumps in $W$ at $\beta_c$ show the strong $1st$ order nature of the transitions for $q > 4$.

Exactly the same picture is seen for the number of triplets $\lambda_3$. For randomly triangulated surfaces the value of $\lambda_3$ at $\beta_c$ decreases with increasing $q$ and is very close to zero for $q = 200$. On a regular lattice the value is on the other hand closer to one, as the cluster boundaries are very complicated and the probability for two neighboring cluster to share a neighbor is very large.

An interesting observation is that on dynamical triangulations both observables seem to reach some kind of plateau value just before $\beta_c$, then start to decrease again. This is more pronounced for $\lambda_3$. Looking at the observables for different lattice sizes indicate that their derivatives are discontinuous in $\beta_c$. A plausible explanation of this is that these quantities approach constant values in the branched polymer phase, slightly larger than the limiting values $W = 1$ or $\lambda_3 = 0$. But as the model enters the pure gravity phase the values drop again, this time due to total magnetization.

## 6 Discussion

We have performed high statistics Monte Carlo simulations of the 10-state and 200-state Potts models coupled to quantum gravity. In a statistical mechanics notation this amounts to approximating the annealed average over triangulations of spherical topology. For both systems we observed a phase transition to a magnetized phase for sufficiently large $\beta$, but the simulations suggest that the two systems belong to different universality classes. The transition for $q = 200$ is compatible with a third order transition, while data clearly favor a second order transition for $q = 10$.

For the 200-state Potts model the dynamics at the transition agrees well the mean field calculations at $q = \infty$. Just below the transition we observe many baby universes which are magnetized individually while the total magnetization is zero. Above the transition the spins of the individual baby universes are aligned. The data are not incompatible with the existence of a region II, below the critical point $\beta_{c_2}$ as suggested in fig.1. From the measurements of $W$, $\lambda_3$ and $\gamma_s$ it seems that there may be a small region below $\beta_{c_2}$ where there are many magnetized baby universe and where $\gamma_s \approx 1/2$. This size of this region could well be comparable with the estimate $\beta_{c_2} - \beta_{c_1} \approx 0.1 \pm 0.05$ mentioned in Sec. 2.

Unfortunately we have not yet been able to locate the phase transition at $\beta_{c_1}$. *If* there is a plateau where $\gamma_s = 1/2$ there *has* to be a second transition since it can been shown from general arguments that $\gamma_s = 1/(n+1)$ if $\gamma_s > 0$, i.e. $\gamma_s$ cannot change continuously to the value $-1/2$ for small $\beta$.

Maybe the most interesting result is that the critical behavior of the 10-state Potts model coupled to two-dimensional quantum gravity seems identical to that of the 4-state Potts model. The 4-state Potts model has central charge $c = 1$, and the result indicates that there exists a whole family of $q$-state Potts models (probably up to $q_c > 10$ of fig.1) which on a regular lattice have first order transitions and no relation to central charge on a regular lattice, but which nevertheless after coupling to gravity lead to the same theory as a $c = 1$ conformal field theory and therefore should allow for a continuum interpretation, but in the case of the Potts models with $4 < q < q_c$ *only after coupling to gravity*. Clearly this calls for a theoretical understanding, and it hints that the $q$-state Potts model coupled to two-dimensional quantum gravity might be exactly solvable.